\documentclass[11pt]{article}
\usepackage{moriond,epsfig}

\def\Journal#1#2#3#4{{#1} {\bf #2}, #3 (#4)}

\def\PLB{{\em Phys. Lett.}  B}
\def\PRL{\em Phys. Rev. Lett.}
\def\PRD{{\em Phys. Rev.} D}

\def\EPJ{{\em Eur. Phys. J} C}

\def\D0{D\O}                            
\def\pt{P_T}                            
\def\et{E_T}                            
\def\gev{\rm \; GeV}                       
\def\mev{\rm \; MeV}                       
\def\ifb{\rm fb$^{-1} \;$}                  
\def\schiH{{\widetilde \chi}^0_2}
\def\schiL{{\widetilde \chi}^0_1}
\def\schiPM{{\widetilde \chi}^\pm_1}
\def\stau{\widetilde \tau_1}             
\def\smu{\widetilde \mu_R}             

\def\met{\mbox{${\hbox{$E$\kern-0.6em\lower-.1ex\hbox{/}}}_T$}} 
\def\mpt{\mbox{${\hbox{$P$\kern-0.6em\lower-.1ex\hbox{/}}}_T$}} 

\def\be{\begin{equation}}
\def\ee{\end{equation}}
\def\bea{\begin{eqnarray}}
\def\bct{\begin{center}}
\def\ect{\end{center}}
\def\eea{\end{eqnarray}}

\begin{document}
\vspace*{4cm}
\title{LEPTON FLAVOR VIOLATION AT LHC}

\author{N. G. UNEL }

\address{Department of Physics, CERN,  Geneva, Switzerland; and\\
Department of Physics, University of California at Irvine, California ,USA.}

\maketitle\abstracts{
Lepton flavor violation (LFV) within the realm of the Standard Model is forbidden. However
recent neutrino experiments strongly suggest neutrino oscillations, giving way to LFV. Beyond SM
theories, such as supersymmetry and supergravity also allow LFV. This note reviews the possibility
of observing LFV signal in the two general purpose LHC experiments: ATLAS and CMS. It is shown that using the initial LHC luminosity, in about a year, either a discovery can be made or the current LFV
limits can be enhanced by an order of magnitude.
}

\section{Introduction}
As it stands today, the Standard Model (SM) does not allow mixing between different lepton families.
However recent results from accelerator, reactor, atmospheric and solar neutrino experiments\cite{neutrinoresults}, point consistently to neutrino oscillations which allow the non-conservation of the lepton flavor at one loop level.  If the generation of the neutrino mass terms, necessary for the oscillation, is done via the introduction $\nu_R$, another SM fine tuning  problem occurs. Why the Yukawa couplings of the same weak isodoublet should be so different for its upper and lower components? As an answer, see-saw mechanism introduces a very large Majorana mass term ($M_R$) compared to Dirac mass terms ($M_D$) of charged fermions: $m_\nu \approx M^2_D / M_R$. 
Post-SM theories such as Supersymmetry (SUSY) try to address this mass generation problem in a more generic way to avoid fine tuning. Supersymmetric models, which would also cure the well known hierarchy and naturalness problems of the SM, could shed light to the mass hierarchy of the proposed three neutrino oscillation scheme and naturally contain   Lepton Flavor Violation (LFV).
Some of the current limits from the particle data group\cite{pdg04} on the LFV processes are:
\bea
\mbox{BR: } \tau \rightarrow \mu \gamma & < & 1.1 \times 10 ^{-6} \nonumber \\
\mbox{BR: } \tau \rightarrow e \gamma & < & 2.7 \times 10 ^{-6} \nonumber \\
\mbox{BR: } \tau \rightarrow \mu \mu \mu & < & 1.9 \times 10 ^{-6} \nonumber \\
\mbox{BR: } Z \rightarrow \mu \gamma & < & 1.5 \times 10 ^{-5} 
\eea 
The existing $b$ factories\cite{bfactory} have recently published results improving these limits by an order of magnitude:
\bea
\mbox{BR: } \tau \rightarrow \mu \gamma & < & 2.0 \times 10 ^{-7} \nonumber \\
\mbox{BR: } \tau \rightarrow \mu \mu \mu & < & 9.0 \times 10 ^{-8}  
\eea
The forthcoming Large Hadron Collider (LHC) will provide enough luminosity to study the  LFV  and if observed, possibly understand its origins.  Around late 2007, the LHC accelerator  will start colliding proton beams at a center of mass energy of 14 TeV. The two general purpose detectors ATLAS and CMS will be ready by that time to profit from the initial low luminosity of $1.2\times10^{33} $cm$^{-2}$s$^{-1}$ which is expected to yield about 10 fb$^{-1}$ in one data taking year. The studies presented in this note, some of which assume  low luminosity justifying the fast simulation techniques and not considering the detector effects such as pile-up, are from these two collaborations\cite{atlas-cms}.

\section{LFV with SM particles}\label{sec:smlfv}
Low luminosity LHC will produce a large number of $W$s, $Z$s and $b$-mesons. The decays of these particles would be a good source of $\tau$s which would yield muons as final state particles.  As will be shown in the following sub-sections, these processes allow a detailed study of LFV. The LFV processes involving electrons as final state particles\cite{elfvSUSY}, although very interesting, are considered to be impractical, since the detector will be flooded with electrons and photons coming from other processes.

\subsection{$\tau \rightarrow \mu \gamma$ channel}
This channel has been considered by both ATLAS and CMS collaborations using Ws as a $\tau$ source. The LFV decay of $\tau$ would give $\mu$ and $\gamma$ as the final state particles (FSP). The background to such a process would be the final state radiation (FSR) and the radiative production of $W$s: $W + \gamma \rightarrow \mu \nu_\mu + \gamma$ and $W \rightarrow \tau \nu_\tau \rightarrow \mu \nu_\mu \nu_\tau + \gamma$.

In the work from ATLAS collaboration \cite{taumugamma-atlas}, the upper limit  of BR:$\tau  \rightarrow \mu \gamma < 10^{-6}$ has been implemented into Pythia \cite{pythia} and  the FSP photons were treated in Photos \cite{photos}. The ATLAS fast simulation tool, ATLFast \cite{atlfast} has been used
to account for the detector effects. The generator level cuts were:
\bea
|\eta|_{\gamma , \mu} &<& 2.5  \nonumber \\
\pt (\mu) &>& 6 \gev  \nonumber \\
\et (\gamma) &>& 20 \gev    \quad . 
\eea
The analysis level cuts were separation of the FSP and muon momentum selection:
\bea
\Delta R_{\gamma , \mu} & >& 0.08   \nonumber \\
 20 \gev > \pt (\mu) &>& 6  \gev  \quad . 
\eea
This simple analysis shows that for low luminosity of 10 fb$^{-1}$, one would expect about 17 background events and less than 8 signal events per running year around the $\tau$ invariant mass.\\

A similar study \cite{taumugamma-cms} was also done by CMS using the full detector simulation and reconstruction tools. The L1 trigger was based on either an energetic photon $\et > 25 \gev$ or an energetic muon $\pt > 20 \gev$ or 
both on a photon and a muon  $\et(\gamma) > 25 \gev  \hbox{ and } \pt(\mu) > 5 \gev$. The event selection around the invariant mass of $\tau \pm 60 \mev$ given by the cuts below yielded a selection efficiency of about 6 percent :
\bea
E(\gamma) & > &18 \gev \nonumber \\
\met & > & 20 \gev \nonumber \\
\pt (\mu) & > & 20 \gev \nonumber \\
0.16 > \Delta R_{\gamma , \mu} & >& 0.07  \nonumber \\
\sigma d_\mu &>& 2 \quad ,
\eea
where $\sigma d_\mu$  is the significance of the $\mu$ impact parameter. The CMS results were very similar to those of ATLAS in this channel. In the case of absence of LFV signal, 
90\% confidence level on the BR can be set after
one year of low luminosity run as  BR: $\tau \rightarrow \mu \gamma \leq 10^{-6}$ and  after one year of nominal luminosity run as  BR: $\tau \rightarrow \mu \gamma \leq 3\times 10^{-7}$.\\

Using $Z$ as a $\tau$ source is an interesting idea and has been pursued \cite{taumugamma-atlas-zz} in ATLAS collaboration. The goal is to use one $\tau$ to tag the event and the other one to investigate the LFV signal. Although the signal is smaller by a factor of 10 compared to the $W$ channel, the selection efficiency is about 14 percent with simple cuts:
\bea
cos (\theta_{\parallel}) & < & 0.5 \; (M_Z) \nonumber \\
N_{\rm jet} & < & 2 \nonumber \\
\mpt & < & 60 \gev   \nonumber \\
\pt (\mu) & >& 6 \gev \nonumber \\
\et (\gamma) & >& 15 \gev  
\eea

In addition to the FSR associated $t \overline{t}$ and $b\overline{b}$, the investigated background processes originate from the muon and tau decay of W and Z:
$W \rightarrow \mu \nu_\mu + \gamma$ 
$W \rightarrow \tau \nu_\tau + \gamma \rightarrow \mu \nu_\mu  \nu_\tau + \gamma $ 
$Z \rightarrow \mu \mu + \gamma$
$Z \rightarrow \tau \tau + \gamma \rightarrow \mu \nu_\mu  \nu_\tau \tau + \gamma \; .$
Using Pythia as event generator and hadronizer, Photos for photon development and ATLFast  for fast detector simulator, this analysis shows that, in an invariant mass range from 1.6 to 2.0 $\gev$, one could obtain about 10 signal events over a background of approximately 12 events per year of low luminosity run yielding a significance of $S/ \sqrt B$ =2.9. In case of signal absence, one gets an upper limit of $2\times 10^{-6}$ on the branching ratio.

\subsection{$\tau \rightarrow \mu \mu \mu$ channel }
Another LFV process studied by CMS starting from Pythia event generator and using  full simulation and reconstruction, is on the three muon decay of the tau lepton\cite{mumumu}. This work also contains a comparison between fast and full simulations showing the pessimistic output from the fast simulation.  $W$, $Z$ and $b$ mesons have been considered separately as  tau sources, with elaborate cuts on both kinematic variables and event topology: 
\bea
\hbox{source} & \hbox{cut level}& \hbox{cut} \nonumber \\
W/Z/B &L1:& \mbox{one}\; \mu \;, \pt (\mu) >  20 \gev \; or\; \mbox{two}\; \mu \;,  \pt (\mu _1) >  8\gev ,\;  \pt (\mu _2) > 5 \gev \nonumber \\
W/Z/B &L3:& \mbox{tree isolated} \; \mu \nonumber \\
W/Z/B &L3:& M_{\hbox{inv}} = 1.777 \pm  0.025 \gev \nonumber \\
W &L3:&  \met > 20  \gev \nonumber \\
Z &L3:&  \pt (\mu) > 20  \gev \nonumber \\
Z &L3:&  M_{\hbox{inv}} (3\mu + \tau_{\hbox{jet}} + \met) > 70  \gev \nonumber \\
B &L3:&  \hbox{2 b-jets in opposite direction one containing 3 $\mu$s  .} 
\eea
The main backgrounds are from $b \overline{b}$ and $c \overline{c}$ production.
 Dedicated analyses have been performed  for the three tau sources trying to optimize the signal significance.
Table \ref{tab:mumumu} gives a brief summary of expected signal and background events per year of low luminosity running. The LFV branching ratio  was taken to be $1.9\times 10^{-6}$.
\begin{table}[h]
\caption{For $\tau \rightarrow \mu \mu \mu$, expected number of LFV signal and background events  per year of low luminosity run. The assumed branching ratio was $1.9\times 10^{-6}$.\label{tab:mumumu}}
\vspace{0.4cm}
\begin{center} \begin{tabular}{|c|c|c|}
\hline
Source & Signal events & Background events \\ \hline
B & 10 $\pm$ 1 & 1 $\pm$ 1 \\ \hline
W  & 46 $\pm$ 2 & 0.8 $\pm$ 1  \\ \hline
Z & 4 $\pm$ 1 & 0.6 $\pm$ 1 \\ \hline
\end{tabular} \end{center}
\end{table}
The invariant mass reconstruction for the most promising source ($W$ channel) is shown in figure \ref{fig:mumumu} as compared to the small SM background which also contains detector effects such as mistags. Using only this channel, in the absence of LFV signal, a 95\% confidence level limit can be set in one year of low luminosity run as BR$ \leq 7\times 10^{-8}$ and as BR$ \leq 3.8\times 10^{-8}$ after three years of low luminosity run.  Although the  Z and B channels give worse results compared to $W$ channel, in case of no-signal, they independently allow reducing the BR limit to  $3.4\times 10^{-7}$ (for the Z channel) and to $2.1\times 10^{-7}$ (for the B channel) with 30 \ifb of data. 
\begin{figure} \bct
\psfig{figure=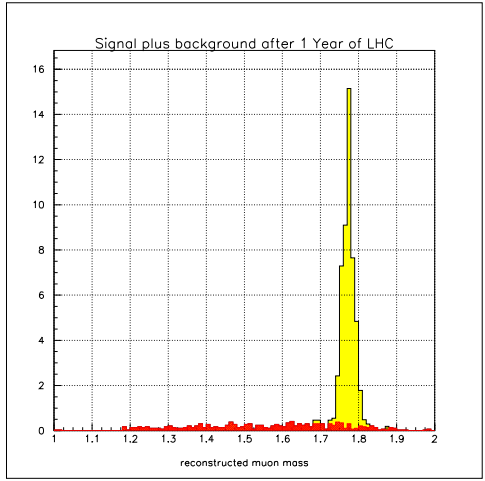,height=2.5in}
\caption{Invariant mass reconstruction from $\tau \rightarrow \mu \mu \mu$ in CMS . $\tau$s originate from W decays, the SM background is the darker shaded area.
\label{fig:mumumu}} \ect
\end{figure}

\section{LFV involving non-SM particles}
Post-SM theories encompass naturally LFV and try to explain some of the (otherwise fine tuned) properties of the SM. The  supersymmetric models can explain naturally neutrino  mixings and mass spectra. For example, in R-parity-violating SUSY\cite{rpvSUSY} there is only one tree level neutrino mass. The loop corrections bring in another mass term, thus explain naturally the experimentally reported neutrino mass difference hierarchy.

\subsection{A scenario with Supersymmetry}

\begin{figure} \bct
\psfig{figure=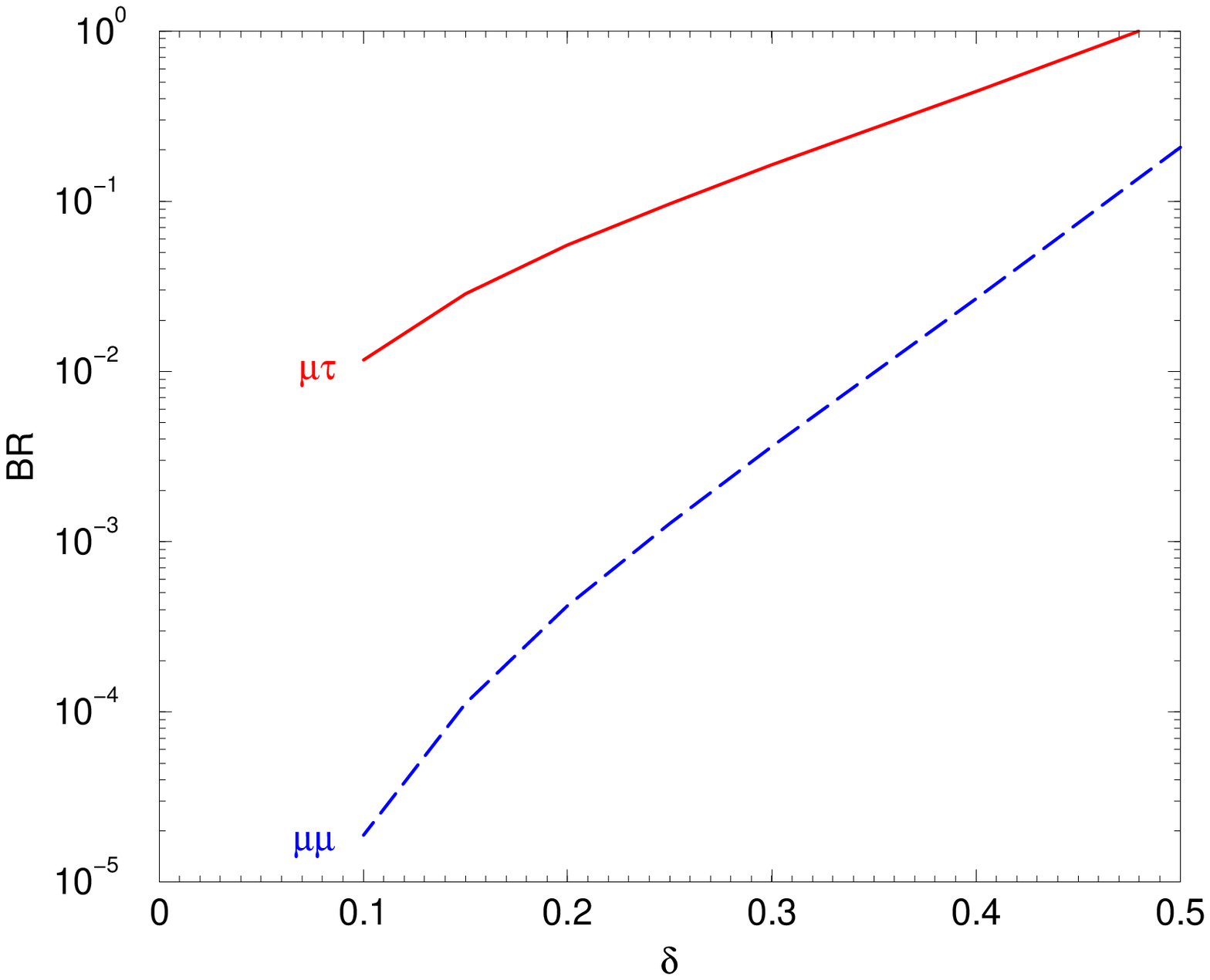,height=1.75in}
\psfig{figure=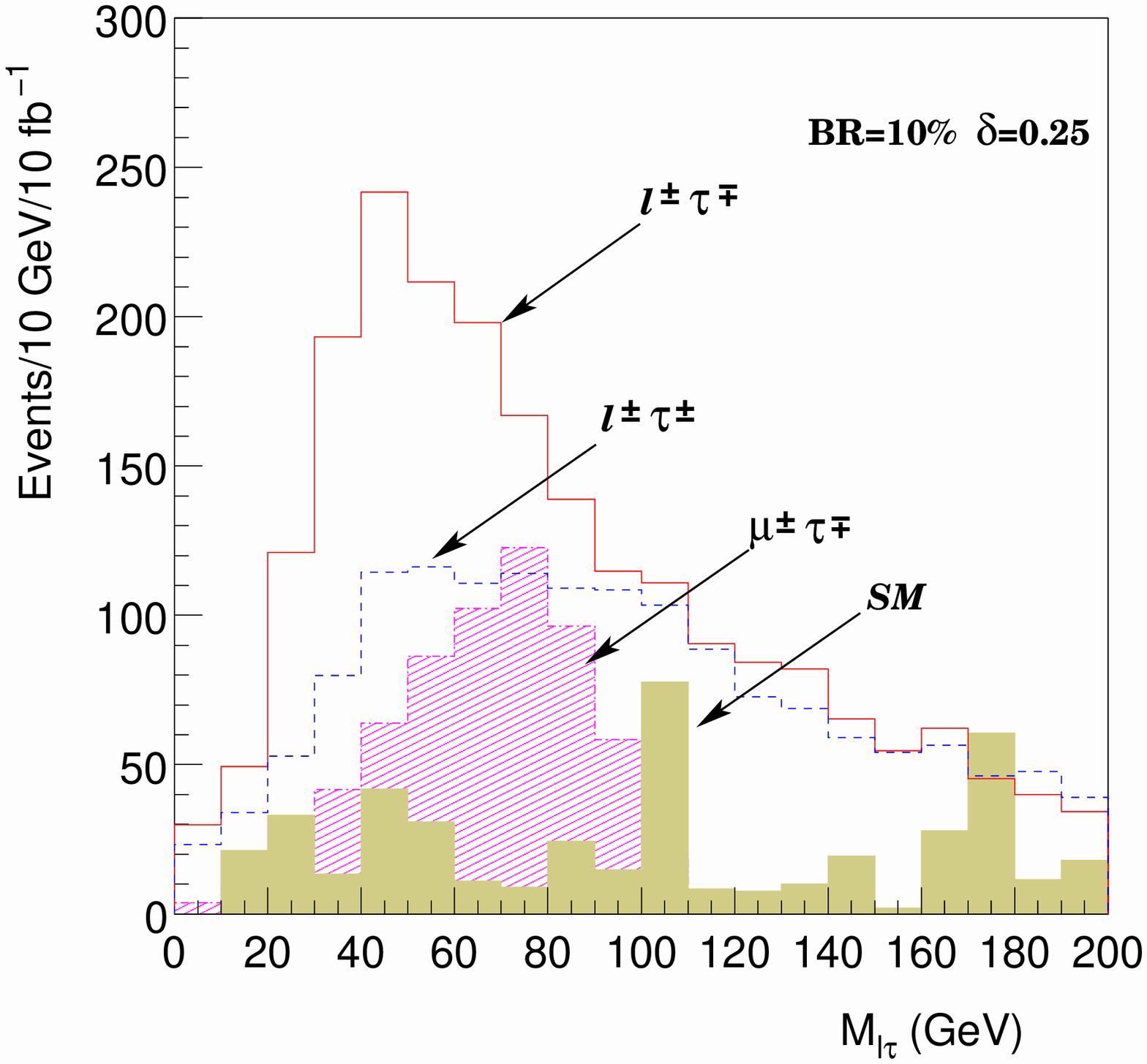,height=1.92in}
\psfig{figure=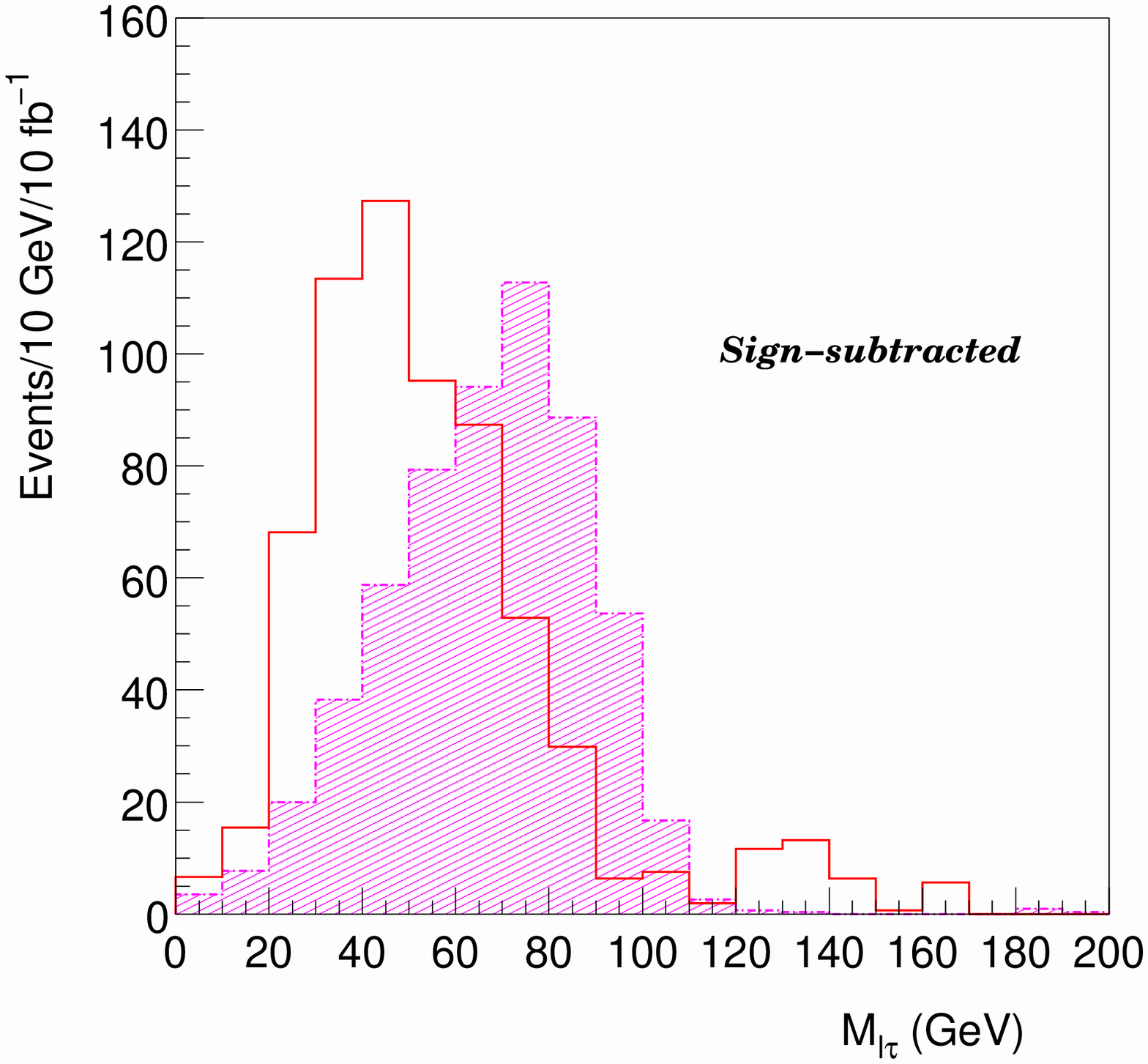,height=1.92in}
\caption {Left to right: a) Branching ratios for 1(solid) and 2(dashed) times LFV decay of neutralino in SUGRA as a function of LFV strength $\delta$. b) Invariant mass distribution for different channels using $\tau - \mu$ final states. c) Separation of LFV (shaded) and non-LFV (solid) signals after sign subtraction. 
\label{fig:SUSY-ab}} \ect
\end{figure}

If the leptons mix, there is no reason why their superpartners should not. Moreover in a supersymmetric theory there is no constraint on the slepton mixing angle. Therefore, the implications of such a model might be large enough to be detected in LHC experiments. A study in ATLAS collaboration investigated the LFV neutralino decays by considering realistic detector effects \cite{SUSY-lfv}. The off-diagonal element of the 6x6 slepton mixing matrix denoted as  $M^2_{\mu\tau}$ is responsible for the $\stau - \smu$ mixing leading to LFV in the $\tau-\mu$ sector. The LFV strength ($\delta$) is the ratio of this quantity to the lepton mass eigenvalue, $\delta \equiv M^2_{\mu\tau} / M^2_L$. If $\delta$ is zero, i.e. no LFV, the heavy neutralino decays mostly to slepton-lepton pairs. \\
The SUGRA point 5 parameters used in this study (except tan$\beta=10$ to have a Higgs mass around 113 $\gev$ to be consistent with LEP II limits) give the branching ratios as,
 BR: $\schiH \rightarrow \stau \tau = 66\%$, BR: $\schiH \rightarrow \smu \mu = 12\%$ and BR: $\schiH \rightarrow {\tilde e}_R e = 12\%$ where $\stau$ and right sleptons are the lightest of the relevant slepton pairs. If $\delta$ is not zero, then there is $\stau - \smu$ mixing which allows other channels such as
 $\schiH \rightarrow \stau \mu \rightarrow \schiL \tau \mu$ and  $\schiH \rightarrow \stau \mu \rightarrow \schiL \mu \mu$ to take place. LFV occurs once in the former decay and twice in the latter. The branching ratios for these two new decay modes as a function of LFV parameter $\delta$ are given in figure \ref{fig:SUSY-ab}a. If one wants to study the LFV signal with $\tau \mu$ final states, the event signature becomes
 missing energy from light neutralino, an isolated muon and multiple jets (one from $\tau$ hadronic decay). The event selection cuts are given as:
 \bea
 N_{\hbox{jet}} & \geq& 4, \; P_T(j1) > 100 \gev \; \pt (j2,j3,j4) > 50 \gev \nonumber \\
 M_{\hbox{eff}} & \equiv & \met + \Sigma \pt > 800 \gev \nonumber \\
 \met & >& 0.2 \times M_{\hbox{eff}} \nonumber \\
 | \eta | & < & 2.5 \nonumber \\
 R & < & 0.4 
 \eea
The SM background to this process is negligibly small. The major contribution comes from the 
tau and muon final state particles of the non-LFV  neutralino decays, $\schiH \rightarrow \tau^\pm \tau^\mp \schiL \rightarrow \mu^\pm \tau^\mp \nu_\mu \nu_\tau \schiL$ and of the two independent chargino decays $\schiPM  \rightarrow \tau^\pm \schiL $. For signal and background processes mentioned, the reconstructed lepton tau invariant masses  are given in figure \ref{fig:SUSY-ab}b. To extract the LFV signal, events with only opposite lepton-tau charges are selected. The result of this sign subtraction is  given in figure \ref{fig:SUSY-ab}c, showing that LFV events have a harder invariant mass distribution. In order to recover them, the excess in $\tau-\mu$ channel compared to $\tau-e$ channel is used (denoted as flavor subtraction). The experimentally interesting quantity is this difference, expected to be zero in case of flavor conservation:
\be
E \equiv N(\mu^\pm \tau^\mp) - N(e^\pm \tau^\mp)
\ee
For one year of low luminosity run and LFV branching ratio of 0.1 $(\delta=0.25)$, we expect $E = 476 \pm 39$.  If at 30\ifb, 5$\sigma$ signal is observed, the value of $\delta$ can be extracted as $\delta = 0.1$ or BR=0.023. If no signal is observed at all in 10\ifb data, the limit obtained on
the branching ratio is better than that from B-factories,  BR$< 10^{-9}$.
\subsection{A scenario with 2HDM}
\begin{figure} \bct 
\psfig{figure=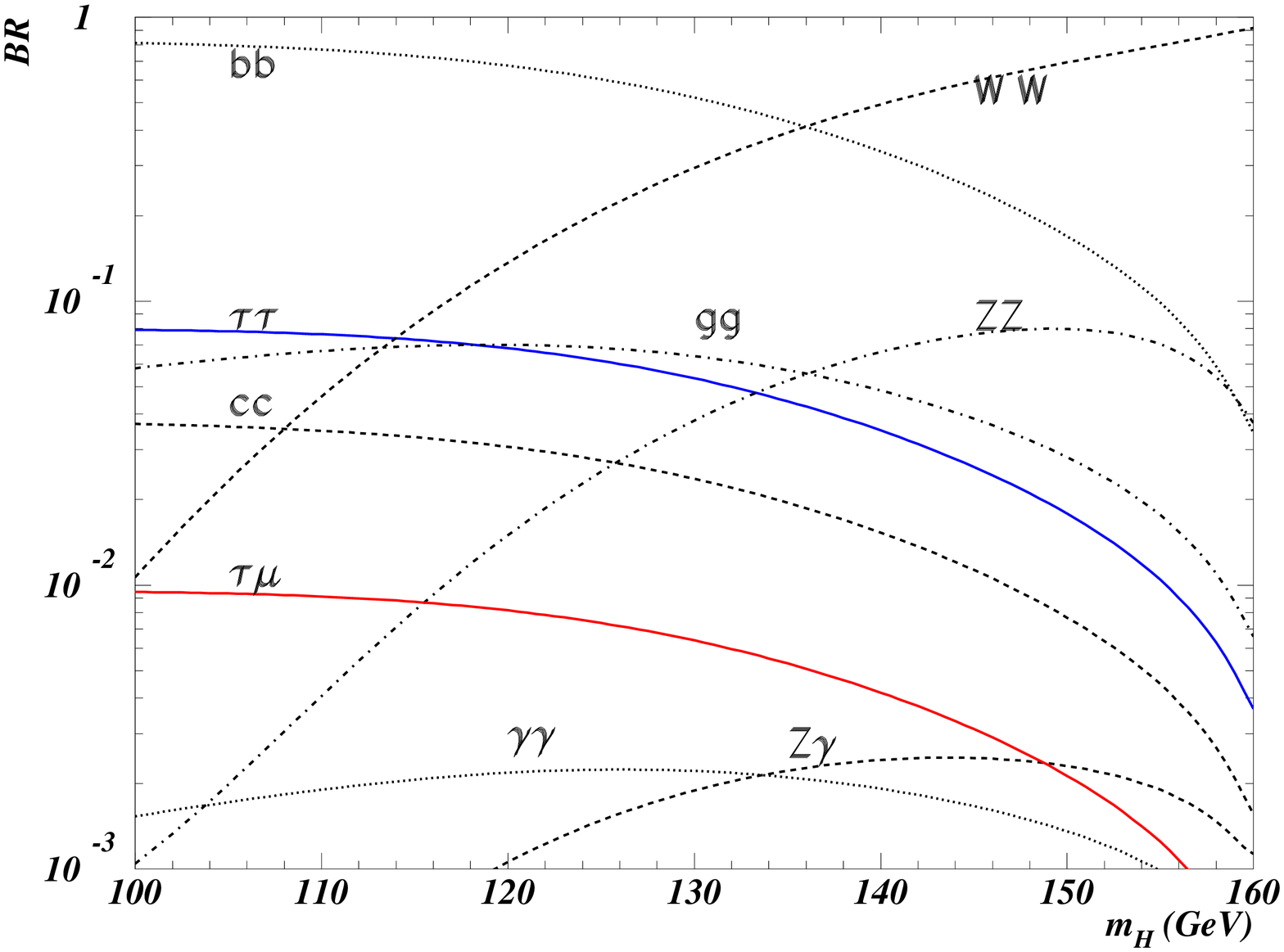,height=1.74in}
\psfig{figure=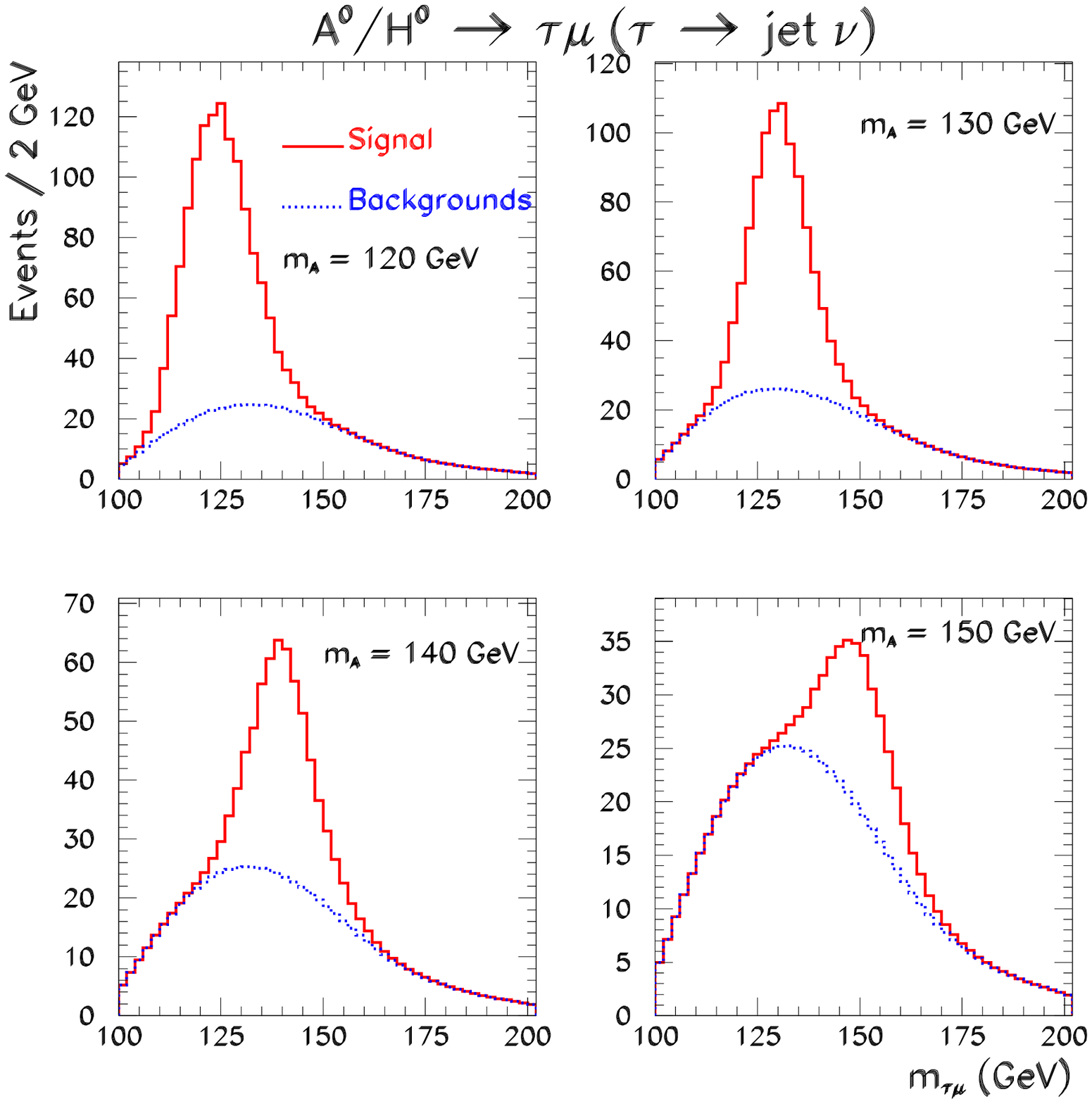,height=1.92in}
\psfig{figure=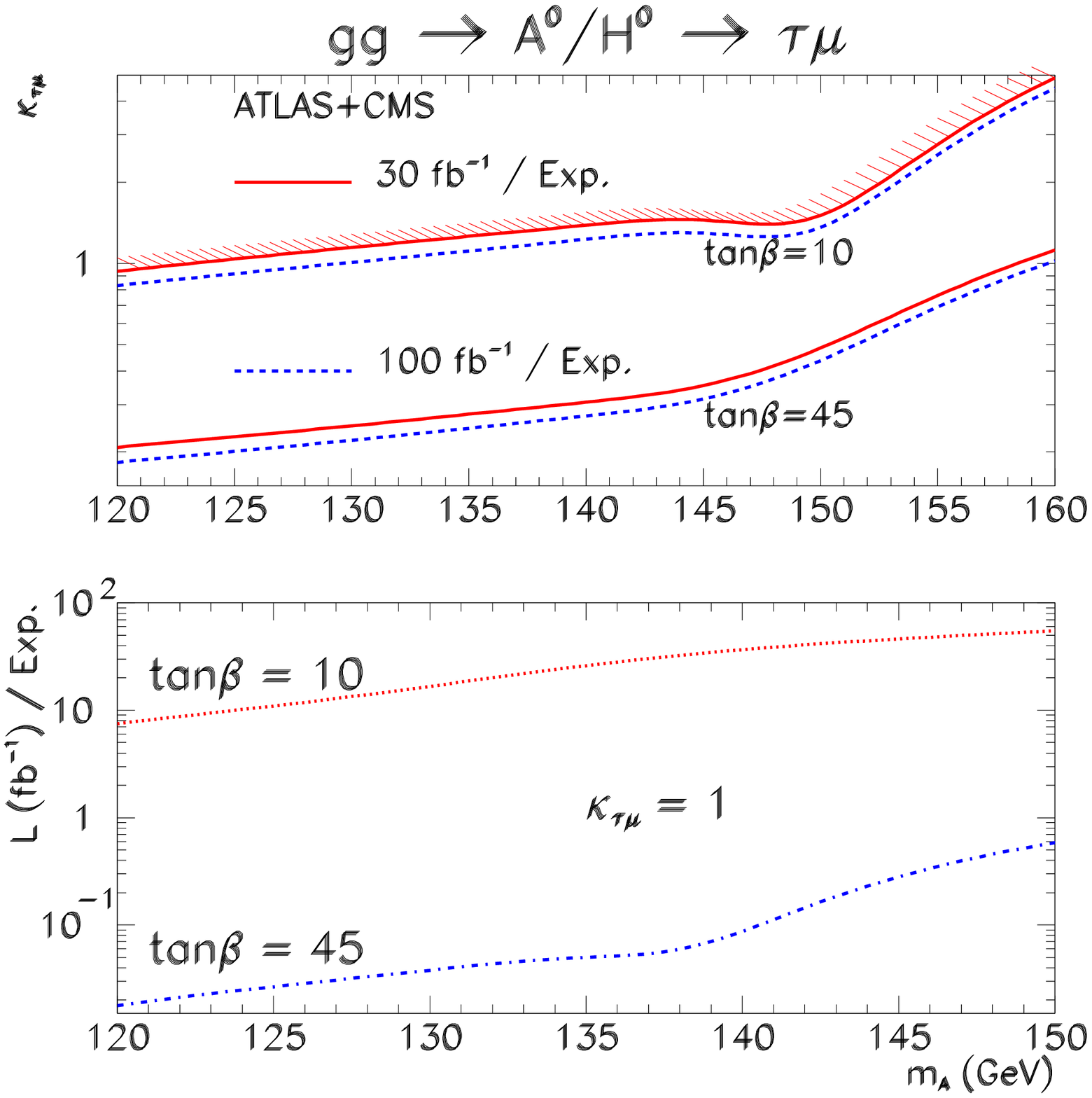,height=1.92in}
\caption {Left to right: a) The branching ratios of the neutral Higgses in 2HDM. b) Signal and Signal + Background events using the LFV neutral Higgs decay for its various masses. c) LFV coverage as a function of neutral Higgs mass for signal (upper) and null (lower) hypotheses .
\label{fig:2hdm-sb}} \ect
\end{figure}
In the models where an additional Higgs doublet is introduced (two Higgs doublet model - 2HDM) to extend the SM, the coupling of Higgs particles to the fermions, defines its type. In type I , up and down types couple to the same Higgs, whereas in type II to a different one. If these conditions are not imposed by hand, type III model is obtained where both fermions couple to both Higgses. Since the diagonalization of the mass matrix is not necessarily the same as of the coupling matrix,  type III 2HDM contains LFV at the tree level. A study\cite{2hdm-lfv} by ATLAS  investigated this possibility starting from neutral Higgses originating from gluon fusion and decaying into tau-muon pairs using fast detector simulation. 

The figure \ref{fig:2hdm-sb}a shows the branching ratios for the possible decays of the additional neutral Higgs as a function of its mass. For the LFV tau-muon channel, the coupling parameter $\kappa_{\tau \mu}$ is taken to be one. In this analysis both hadronic jet decay and muon decays of $\tau$ is considered.  The backgrounds come from $W^\pm Z$ , $W^+ W^-$, $t \overline{t}$, $Z$, $W^\pm +jets$ and $A^0/H^0$ decays involving tau-muon pairs and their combined cross section is orders of magnitude larger than the signal. Separate analyses were developed for hadronic and leptonic tau decay identification based on good tracking,  $\met$ and $b$-jet identification in the barrel region. 
With a $\tau$-jet identification of 30\%, a selection of only 1-prong $\tau$ decays, and  cuts on jet cone angles, the SM background can be reduced to a manageable level. The backgrounds originating from  $\tau \overline{\tau}$ decays of $A^0/H^0$, can be very large depending on the model parameters. The rejection of these is possible with good $\mpt$ resolution and $\tau$-jet identification. The resulting number of signal and signal+background events for 30 \ifb of data are shown in figure \ref{fig:2hdm-sb}b for tan$\beta$=45, $\kappa_{\tau \mu}$=1 and various neutral Higgs masses. Up to 150 GeV it is possible to obtain a large significance for the LFV signal; around 160 GeV the introduction of the $W^+ W^-$ decay mode makes the ${\tau \mu}$ channel unmeasurable.  

When both ATLAS and CMS data are combined, the 5$\sigma$ discovery plane in the upper part of the figure \ref{fig:2hdm-sb}c is obtained. Higher values of tan $\beta$ allow reaching values of $\kappa_{\tau \mu}$ as low as 0.18 for either three years of low luminosity run (solid lines) or one year of nominal luminosity run (dashed lines). If no LFV signal is found,  the necessary luminosity for 95\% 
confidence  level exclusion as a function of the Higgs mass is shown in the lower plot of figure \ref{fig:2hdm-sb}c for two values of tan $\beta$.  With 30 \ifb, a 95 \% CL upper-limit on the branching ratio can be set up to a Higgs mass of 150 GeV. 
\section{Conclusions}
Low luminosity LHC, which will start towards the end of 2007, will be an abundant source of $\tau$'s that can be used to search for LFV signals. The analyses done in both ATLAS and CMS collaborations show that if Nature gives  large enough branching ratios  in the access range of the LHC, the experimental sensitivity and the background elimination in both experiments are good enough to observe LFV signal within few years of low luminosity runs. Some post-SM theories such as SUSY, justify  this expectation by enhancing the signal expectations. If no signal at all is observed, after three years of low luminosity run, the limits on LFV from LHC experiments can be an order of magnitude more stringent than current ones.

\section*{Acknowledgments}
The author would like to thank  G. Azuelos of the ATLAS experiment for fruitful discussions 
and P. Sphicas of the CMS experiment for providing results of the previous studies. He also expresses his gratitude to the organizers of the Moriond Conference for their warm hospitality and to L. Mapelli for lending some of the ATLAS trigger and data acquisition manpower to physics studies.

\end{document}
